\RequirePackage{amsmath}
\documentclass[epj]{svjour}
 
\usepackage{graphics}
\usepackage{amssymb}
\usepackage{amsfonts}
\usepackage{amsbsy}
\usepackage{float}
\usepackage{pstricks}
\usepackage{graphicx}
\usepackage{multirow}
\usepackage{hyperref}
\usepackage[]{subfigure}
\usepackage{epsfig}
\usepackage{amssymb}
\usepackage{graphicx}
\usepackage{gensymb}
\usepackage{amsmath}
\usepackage{relsize}
\usepackage{mathtools}
\usepackage{alphalph}
\usepackage{hyperref}
\usepackage{blindtext}

\newcommand{\be}{\begin{equation}}
\newcommand{\ee}{\end{equation}}
\newcommand{\ba}{\begin{eqnarray}}
\newcommand{\ea}{\end{eqnarray}}

\def\beq{\begin{equation}}
\def\eeq{\end{equation}}
\def\beqq{\begin{eqnarray}}
\def\eeqq{\end{eqnarray}}

\newcommand{\bdm}{\begin{displaymath}}
\newcommand{\edm}{\end{displaymath}}

\begin{document}
\title{\boldmath Spin, torsion and violation of null energy condition in traversable wormholes } 
\author{Elisabetta Di Grezia  \inst{1} \and Emmanuele Battista \inst{1} \and Mattia Manfredonia \inst{1,2} \and Gennaro Miele\inst{1,2}}

\institute{INFN, Sezione di Napoli, Complesso Univ. Monte S. Angelo, I-80126 Napoli, Italy \and Dipartimento di Fisica Ettore Pancini, Universit\`a di Napoli Federico II, Complesso
Univ. di Monte S. Angelo, I-80126 Napoli, Italy.  \\ \email{ebattista@na.infn.it, digrezia@na.infn.it,  manfredonia@na.infn.it, miele@na.infn.it}}

\date{Received: date / Revised version: date}

\abstract{The static spherically symmetric traversable wormholes are analysed in the Einstein-Cartan theory of gravitation. In particular, we computed the torsion tensor for matter fields with different spin $S=0, 1/2, 1,  3/2$. Interestingly, only for certain values of the spin the torsion contribution to Einstein-Cartan field equation allows one to satisfy both flaring-out condition and Null Energy Condition. In this scenario traversable wormholes can be produced by using usual (non-exotic) spinning matter. }

\maketitle

\section{Introduction}

In general relativity there exists a number of physically reasonable restrictions on the stress-energy tensor able to guarantee a level of {\it physical acceptability} of field solutions \cite{Hawking_large,Horowitz}. However, such so-called energy conditions express only possible criteria that have a certain degree of arbitrariness. 
In this paper we focus on the so-called   {\em Null Energy Condition} (NEC), i.e. $\forall N^a$ such that $N^aN_a =0$, one requires
$T_{ab}\, N^a N^b \geq 0$, with $T_{ab}$  the stress-energy tensor.  NEC results to be the weakest one among the possible energy conditions proposed\footnote{ Indeed, the energy conditions SEC (Strong Energy Condition), NEC, WEC (Weak Energy Condition)\cite{Visser},  and DEC (Dominant Energy Condition)\cite{Visser} satisfy  the following  implication chains:
\begin{equation}
(DEC) \Rightarrow (WEC) \Rightarrow   (NEC) 
\end{equation}
and
\begin{equation}
(SEC) \Rightarrow  (NEC).
\end{equation}
For the definitions of the energy conditions and a detailed discussion about the above implications see for example Ref.s \cite{Visser,Hawking_large}.}. For this reason, one expects that in general a physically reasonable matter should at least satisfy the NEC. Moreover, as well known NEC is explicitly summoned in the Morris-Thorn wormhole
and this feature is typically interpreted as a characteristic that compromises the physical attainability of a material sourcing it, and hence it implicitly suggests that Morris-Thorne wormholes never occur in reality. In this paper we consider  the link between NEC and wormhole in a more general framework.

By adopting a suitable orthonormal frame where $T_{\hat{\mu}\hat{\nu}}={\rm diag}(\rho,p_1,p_2,p_3)$, NEC implies
\begin{equation}
\rho + p_j \geq 0, \; \forall j \in \{1,2,3\}. \label{NEC0}
\end{equation} 

Nevertheless, it is well known that there exists a number of classical and quantum systems capable of violating this constraint \cite{Flanagan-wald,PLB,Visser,Rubakov:2014jja}. In this concern a relevant example is provided by Dirac field in General Relativity.

As already stated, an interesting hypothetical physical system that implies the violation of NEC is the matter sourcing  Lorentzian traversable Morris-Thorne wormholes. These are classical solutions of Einstein equations representing short-cuts between two asymptotically flat region of space time \cite{Morris:1988cz,PLB,Chianese:2017qlx}. In the following we summarize the main features of a traversable Morris-Thorne wormhole.

Let us consider the  most general static and rotationally invariant metric without horizons, which in the Schwarzschild coordinate system $(t,r,\theta ,\phi)$~\cite{Morris:1988cz,Visser,Visser:1992qh} reads
\begin{equation}
ds^2 = -e^{2\Phi(r)} dt^2 +\left(1-\frac{b(r)}{r}\right)^{-1} dr^2+r^2d \Omega ^2,
\label{eq:w_metric}
\end{equation}
where the $b(r)$ and $\Phi(r)$ are the {\it shape} and {\it red-shift} functions, respectively.\\ Although the previous metric has no event horizons, further constraints have to be imposed in order to obtain a traversable wormhole (see Ref.~\cite{Morris:1988cz} for a detailed discussion). Indeed, at any given time $t$, the  bridge-like behavior implies that three-dimensional Euclidean embedding of the manifold equatorial slice have to flare outward as one moves from the upper to the lower universe trough the throat. The above geometrical constraint is known as {\it flaring outward condition} and it implies ~\cite{Morris:1988cz,Visser}
\begin{equation}
b'(r) < \frac{b(r)}{r}\,,
\label{eq:foc}
\end{equation}
near the {\it wormhole  throat}, which is defined as the narrowest region of the wormhole where the radial coordinate takes its minimum value $r_0$~\cite{Morris:1988cz,Visser}. Furthermore, it can be shown  that  $b(r_0)=r_0$~\cite{Morris:1988cz}. Hence in  $r_0$ one gets from Eq.~(\ref{eq:foc})
\begin{equation}
b'(r_0) < 1 \,.
\label{eq:bminorone}
\end{equation}
Through the Einstein equations
\begin{equation}
G_{ij}=k \, T_{ij}
\end{equation}
where $k=8 \pi G$ in natural units,  one translates the geometrical condition \eqref{eq:bminorone} on $G_{ij}$ in a corresponding condition on the components of the stress-energy tensor in an orthonormal frame, namely
\begin{equation}
\tau_0>\rho_0.
\label{exotic-matter}
\end{equation}
The quantities $\rho_0$ and $\tau_0$ represent the values at the throat of total density of mass-energy and the tension per unit area measured in the radial direction, respectively.  Equation \eqref{exotic-matter} indicates that the matter sourcing the traversable wormhole violates NEC in a neighborhood of the throat. A material characterised by such a behaviour is named {\it ``exotic''} \cite{Visser}. This may implicitly suggest that Morris-Thorne wormholes never occur in reality. Interestingly, the above considerations do not apply in general to Einstein Cartan theory\cite{Anchordoqui,Bronnikov,Jawad,Mehdizadeh,Bolokhov:2015qaa,Galiakhmetov,Jawad,Obukhov} where non exotic sources of wormholes are possible. 

In this paper we show how Einstein-Cartan theory allows one to disjoin the flaring out condition from the NEC violation. The result is strictly related the torsion contributions arising in the field equations and hence to the role played by the spin. 

In Einstein-Cartan theory ($U_4$ theory)  \cite{Cartan,Kuchowicz} the space-time has a non trivial Riemannian structure since the antisymmetric part\footnote{Note that the symmetric part usually does not transform as a tensor (Christoffel symbol)} of the connection no longer vanishes.  This new term contributes to the so-called modified torsion tensor  $T_{ijk}$  \cite{Borut} and figures in field equations in addition to the curvature. Interestingly, in this extension of General Relativity the torsion is related to the  intrinsic angular momentum  just as the Ricci tensor is related to the energy-momentum tensor. Since the presence of torsion should be associated to a certain large energy scale \cite{Shapiro} one expects to see its effects at work in the very early universe only \cite{Gasperini,Garcia,Palle}. 

This paper is organized as follows. In Section 2  we briefly review the Einstein-Cartan theory. In Section 3  we investigate the relation between flaring-out condition and NEC violation. In particular we discuss some relevant particle fields \cite{Smalley,Barth} with integer and half-integer spin: a ``Dirac field" $S=1/2$ \cite{Seitz1}, a ``Proca field" $S=1$ \cite{Seitz,Spinosa1} and a ``Rarita-Schwinger field" $S=3/2$ \cite{Spinosa}. Section 4 is dedicated to conclusions and outlooks.

\section{Einstein-Cartan theory}

In this section we briefly review the formalism of Einstein-Cartan-(Sciama-Kibble) theory, sometimes refereed to as $U_4$ theory \cite{Sciama,Kibble:1961ba,Hehl:1976kj}, in order to stress the main concepts and define the notation. 
In this theory both  mass and spin of matter fields couple to the geometry of spacetime. Equivalently, such a model arises by constructing a gauge theory whose (local gauge) group is represented by Poincar\`e group. Within both approaches, we  find that spin couples to torsion of the four dimensional Riemann-Cartan spacetime $U_4$ in the same way as the energy-momentum tensor couples to the metric (see Eqs. (\ref{Hehl-3.21}) and (\ref{Hehl-3.22}) below).

Let us consider a four-dimensional spacetime $X_4$ endowed with a metric $g_{ij}$, where $i,j=0,1,2,3$. The most general metric compatible connection (i.e., the one for which the metricity condition $(\stackrel{\Gamma}{\nabla}_i g)_{jk}=0$ is satisfied\footnote{We denote the covariant derivative carried out through the connection $\Gamma$ with the symbol $\stackrel{\Gamma}{\nabla}$.}) arising in $X_4$ is given by
\begin{equation}
\Gamma^{i}_{\; jk}= \Gamma^{i}_{\; (jk)} + \Gamma^{i}_{\; [jk]}=\left \{ 
\begin{array}{cc}
i \\ jk
\end{array}
\right \}-K_{jk}^{\;\;\;i},
\label{hehl-2.11}
\end{equation}
where $\left \{ 
\begin{array}{cc}
i \\ jk
\end{array}
\right \}$ are the Riemann-Christoffel symbols, whereas
\begin{equation}
\Gamma^{i}_{\; (jk)} = \dfrac{1}{2} \left( \Gamma^{i}_{\; jk}+\Gamma^{i}_{\; kj}\right)=\left \{ 
\begin{array}{cc}
i \\ jk
\end{array}
\right \}+ S^{\;\; i}_{j\;\;k}+S^{\;\;i}_{k\;\;j},
\end{equation}
\begin{equation}
\Gamma^{i}_{\; [jk]} = \dfrac{1}{2} \left( \Gamma^{i}_{\; jk}-\Gamma^{i}_{\; kj}\right)=S_{jk}^{\;\;\;i},
\label{torsion_def}
\end{equation}
and hence
\begin{equation}
K_{jk}^{\;\;\;\,i}=-S_{jk}^{\;\;\; \,i}+S_{k\;\;\;j}^{\;\;\,i}-S^{i}_{\;jk}=-K_{j\;\;\;k}^{\;\;\,i}.
\end{equation}
The quantities $K_{jk}^{\;\;\;\,i}$ and $S_{jk}^{\;\;\;i}$ have 24 independent components and are called contorsion and Cartan torsion tensors, respectively. The $K_{jk}^{\;\;\;\,i}$, representing the non-Riemannian part of the connection,  depend both on metric and on torsion. A spacetime $X_4$ endowed with the connection (\ref{hehl-2.11}) is called Riemann-Cartan spacetime $U_4$. Note that for a vanishing torsion one recovers the usual Riemannian spacetime.

Let us consider in $U_4$  a matter field which couples to gravity. The total action reads simply as
\begin{eqnarray}
W_{{\rm tot}}&=&W_{{\rm m}}+W_{{\rm g}} =  \int {\rm}d^4 x  \left[\mathcal{L}\left(\psi, \partial \psi,g, \partial g, S\right) \right.\nonumber\\
&+& \left. \dfrac{\sqrt{-g}}{2k} R\left(g,\partial g,S,\partial S \right) \right]
\label{Hehl-3.15}
\end{eqnarray}
where $R$ is the Ricci scalar.
If we vary (\ref{Hehl-3.15}) with respect to $\psi$ we obtain the matter equations
\begin{equation}
\sqrt{-g} \,\Sigma_i^{\;\, j}= \mathcal{L} \, \delta_i^{\;j}-\dfrac{\partial \mathcal{L}}{\partial (\partial_j \psi)}\stackrel{\Gamma}{\nabla}_i \psi,
\label{canonical-energy-momentum-tensor}
\end{equation}
\begin{equation}
\sqrt{-g}  \,\tau_{k}^{\;\,ji} = \dfrac{\delta \mathcal{L}}{\delta K_{ij}^{\;\;\,k}},
\label{canonical-spin-tensor}
\end{equation}
where on the left we have the canonical energy-momentum tensor and the canonical spin tensor, respectively. Eq.\eqref{canonical-spin-tensor}
expresses the coupling between spin and contorsion. 

By varying the total action (\ref{Hehl-3.15}) with respect to field variables $g_{ij}$ and $S_{ij}^{\;\;\, k}$ we get two sets of field equations
\begin{equation}
\begin{split}
& -\dfrac{\delta \left( \sqrt{-g}\,R\right)}{\delta g_{ij}} = k \sqrt{-g} \, \sigma^{ij}, \\
& -\dfrac{\delta\left(  \sqrt{-g}\,R\right)}{\delta S_{ij}^{\;\; \,k}}= 2 k \sqrt{-g} \, \mu_k^{\;\, ji},
\label{Hehl-3.16}
\end{split}
\end{equation}
where $\sqrt{-g}\, \sigma^{ij} = 2 \dfrac{\delta \mathcal{L}}{\delta g_{ij}}$,$\ \  \sqrt{-g} \, \mu_{k}^{\;\, ji}= \dfrac{\delta \mathcal{L}}{\delta S_{ij}^{\;\; k}}$ 
are the metric energy tensor and the spin potential energy, respectively. The tensors $ \Sigma^{ij},\sigma^{ij}, \tau^{ijk}, \mu^{ijk}$ are related by the following relations:
\begin{equation}
\mu^{ijk}=-\tau^{ijk} + \tau^{jki}-\tau^{kij},
\end{equation}
\begin{equation}
\Sigma^{ij}= \sigma^{ij}-\stackrel{*}{\nabla}_k \mu^{ijk},
\label{Hehl-3.8}
\end{equation}
where $\stackrel{*}{\nabla}_k \equiv \stackrel{\Gamma}{\nabla}_k + 2S_{kl}^{\;\;\,l} $. The last equation represents the decomposition of the asymmetric canonical energy-momentum tensor into a symmetric part provided by the metric energy-momentum tensor and an antisymmetric term involving the spin.  If one considers the canonical tensors $\Sigma^{ij}, \tau^{ijk}$ as the sources of the field equations (\ref{Hehl-3.16}), after some calculations (see Ref. \cite{Hehl:1976kj} for details)  Eq. (\ref{Hehl-3.16}) becomes 
\begin{equation}
G^{ij} = k \,\Sigma^{ij},
\label{Hehl-3.21}
\end{equation} 
\begin{equation}
T^{ijk}=k\, \tau^{ijk}.
\label{Hehl-3.22}
\end{equation}
Equation (\ref{Hehl-3.21}) represents a generalization of Einstein equations, whereas Eq. (\ref{Hehl-3.22}) is an {\it algebraic} relation featuring the coupling between the spin of matter and the torsion of spacetime. 
The  Einstein tensor appearing in Eq. (\ref{Hehl-3.21}) has been defined as usual
\begin{equation}
G_{ij}=R_{ij}-\dfrac{1}{2}g_{ij}R.
\end{equation}
The antisymmetric part of $G_{ij}$ is now non-vanishing due to the modified divergence relation
\begin{equation}
G_{[ij]}=\stackrel{*}{\nabla}_k T_{ij}^{\;\; k}.
\end{equation} 
By employing the algebraic relation (\ref{Hehl-3.22}), it is possible to write Eqs. (\ref{Hehl-3.21}) and (\ref{Hehl-3.22}) as an unique set of field equations given by
\begin{equation} \label{riem_G}
\stackrel{\{\}}{G}_{ij}=k \,\tilde{\sigma}_{ij},
\end{equation}
where $\stackrel{\{\}}{G}_{ij}$ represents the Riemannian part of $G_{ij}$ and
\begin{eqnarray} \label{sigmatilde}
\tilde{\sigma}^{ij}&=&\sigma^{ij} + k \left[-4 \tau^{ik}_{\,\;\;[l}\tau^{jl}_{\;\; \, k]}-2\tau^{ikl} \tau^{j}_{\;\,kl} + \tau^{kli}\tau_{kl}^{\;\;\,j} \right.\nonumber\\
& + & \left.\dfrac{1}{2}g^{ij} \left( 4 \tau_{m \;\; [l}^{\;\;k}  \tau^{ml}_{\;\;\;\;k]} +\tau^{mkl}\tau_{mkl}\right)\right].
\end{eqnarray}

\section{Spin and NEC} \label{Spin and NEC_Section}
The NEC is often required in general relativity to avoid unpleasant behaviours of spacetime \cite{Hawking_large}. At the same time, NEC violations occur in particular (maybe just hypothetical) spacetime configurations. As shown in the introduction and in Refs. \cite{Morris:1988cz,Visser}, the ``{\it flaring-out condition}'' for static spherically symmetric wormholes implies NEC to be violated. For such reason, the matter driving a wormhole is typically denoted as {\it exotic} \cite{Visser}. 

Although the flaring-out condition and the  NEC violation are usually strictly related, they are not equivalent. Indeed, flaring out is a geometrical condition which has to be imposed on the Einstein tensor $G_{i j}$. On the other hand, NEC is a physical constraint applied to the stress-energy tensor $T_{ij}$. In ordinary general relativity these two tensors are the only quantities entering the Einstein's equations $G_{ij}=k T_{ij}$ and this implies the equivalence between geometrical and physical constraints. On the contrary, in Einstein-Cartan theory the torsion terms in \eqref{sigmatilde} disjoins the Riemannian part of the Einstein tensor \eqref{riem_G} from the usual energy-momentum tensor related to $\sigma_{ij}$ by \eqref{Hehl-3.16}. It follows that also NEC and flaring-out are now disjoint. Given a generic null vector field $N_{i}$ (i.e., $g^{ij}N_i N_j=0$), the conditions read as follows:
\begin{equation} \label{NEC_FOUT_1}
\renewcommand{\arraystretch}{2.0}
\begin{dcases}
&\stackrel{\{\}}{G\ ^{ij}}N_iN_j=k\, \tilde{\sigma}^{ij}N_iN_j<0, \;{\rm(flaring - out)},
\\
&\sigma^{ij}N_i N_j>0,  \; {\rm (null \; energy \; condition)},
\end{dcases}\\[2em]
\end{equation}
The purpose of this section is to  investigate if conditions \eqref{NEC_FOUT_1} could be simultaneously verified. Since the spin is coupled with the torsion terms, we will focus our attention on some relevant examples of spinning matter theories with different spin number.

\subsection{Scalar Matter Field $(S=0)$}
We start with the trivial case concerning a scalar matter field $\phi$. Since $\phi$ has no spin, it does not produce any torsion. Indeed, the gauge covariant derivative operator  acts on a scalar field as $D_i \phi=\partial_i \phi$ and hence, according to the minimal coupling procedure, torsion does not couple with matter\footnote{It is worth noting that by employing a non-minimal coupling  scalar fields can source the torsion \cite{Bolokhov:2015qaa}}. Therefore, Eq.\eqref{riem_G} reduces to the usual Einstein field equations. This is the same situation as described in Refs. \cite{Morris:1988cz,Visser}, where the flaring-out condition implies NEC violation, i.e.,
\begin{equation}
\begin{split}
&{\rm (Flaring \; Out)}  \;\;\;\;\;\;\;\;\; {\rm (NEC \; violation)}. \\
& G_{ij}N^iN^j<0 \ \ \ \Leftrightarrow \ \ \ T_{ij}N^iN^j<0 .
\end{split}
\end{equation}
When this occurs, the exoticity of the wormhole source is unavoidable \cite{Morris:1988cz}.

\subsection{Dirac Field $(S=\frac{1}{2})$}

The interacting Dirac field \cite{Hehl:1976kj,Shapiro,Watanabe:2004nt,Brill:1957fx} is described by a Lagrangian function whose covariantized form reads as 
\begin{equation}
\mathcal{L}_{{\rm D}}= \sqrt{-g}\, \frac{i}{2}\, \bar{\psi} \left(\gamma^i D_{i}-m\right) \psi.
\end{equation}
The gauge covariant derivative $D_i$ acts on spinor as 
\begin{equation}
D_i \psi=\left(\partial _i -\frac{i}{4}\,\omega _i ^{\; \alpha \beta}\, \gamma_{\alpha \beta}\right)\psi,
\end{equation}
where
\begin{equation}
\gamma_{\alpha \beta}=\dfrac{1}{2} \left[\gamma_\alpha,\gamma_\beta \right],
\end{equation}
$\gamma_\alpha = e^i_{\;\alpha} \gamma_i$ being the Dirac matrices in flat space. The canonical spin tensor is completely antisymmetric and is given by
 \begin{equation}
  \tau ^{\alpha \beta \gamma}=\frac{1}{4}\bar{\psi} \gamma^{[\alpha}\gamma^{\beta} \gamma^{\gamma ]}  \psi,
\end{equation}  
and hence from Eqs.\eqref{riem_G} and \eqref{sigmatilde} we get
\begin{equation}
\stackrel{\{\}}{G}_{ij}=k\left[ {\sigma}_{(ij)}+\frac{3}{16} g_{ij} ( \bar{\psi}\gamma_5 \gamma_0 \psi)^2\right]. \label{Dirac_G}
\end{equation}
Since the only correction occurring in Eq. \eqref{Dirac_G} is proportional to the metric tensor, it follows that such a term vanishes when one computes $\stackrel{\{\}}{G}_{ij} N^i N^j$. Once again, we obtain the usual relation between NEC and flaring out. Therefore, we conclude that only ``exotic'' Dirac fields could satisfy Einstein's equation for wormholes of type discussed in Refs. \cite{Morris:1988cz,Visser}. 
\subsection{Proca Field $(S=1)$}
The Lagrangian function of a Proca field is represented by
\begin{equation}
\mathcal{L}_{{\rm P}}=-\sqrt{-g}\frac{1}{2}\left(\nabla_{[i} U_{j]} \nabla^{[i} U^{j]} -m^2 U_i U^i\right),
\end{equation}
where $m \neq 0$ is a mass \footnote{If one sets $m=0$ then obtains Maxwell theory. As discussed in \cite{Hehl:1976kj} Maxwell fields can be minimally coupled to torsion only by breaking the gauge symmerty.} and $U_i$ is a Lorentz vector valued zero-form belonging to the (irreducible) vector representation of the Lorentz group \cite{Seitz:1986sc}. Thus, its gauge covariant derivative is given by
\begin{equation}
D_i U_j = \left( \partial_i -\dfrac{1}{2} \omega_i^{\; \alpha \beta}J_{\alpha \beta} \right) U_j,
\end{equation}
where $J_{\alpha \beta}$ are the generators of the Lorentz group in the given representation, i.e.,
\begin{equation}
\left(J^{\alpha \beta}\right)^{\gamma}_{\; \delta}= i \left( \eta^{\alpha \gamma}\delta^{\beta}_{\; \delta}-\eta^{\beta \gamma}\delta^{\alpha}_{\; \delta}\right). 
\end{equation}
The canonical spin tensor is given by
\begin{eqnarray} \label{tau_Proca}
\tau_{ijk}&=&\frac{1}{2} \left( U_k \nabla_{[j} U_{i]}-U_j \nabla_{[k} U_{i]}\right) \nonumber\\
&=& \frac{1}{2} \left(U_k F_{ji}-U_{j}F_{ki} \right),
\end{eqnarray}
where we have introduced the tensor $F_{ij} \equiv\nabla_{[i}U_{j]}$. By using Eq. \eqref{tau_Proca} jointly  with Eq. \eqref{sigmatilde} we achieve the following expression:
\begin{eqnarray}
\tilde{\sigma}^{ij}&=& \sigma^{ij} +\frac{k}{2} \left[-U_lU_kF^{ki}F^{lj} \right.\nonumber\\
&+& \left. \frac{1}{2}g^{ij}\left(U^mU^kF^l_{\;k}F_{lm}\right)\right].
\end{eqnarray}
Unlike the Dirac case, the above equation involves a correction term which is not proportional to the metric. Therefore, non vanishing torsion terms appear in the flaring-out condition, i.e.,
\begin{equation}
\tilde{\sigma}^{ij} N_i N_j=\sigma^{ij}N_iN_j-\frac{k}{2}\left(U_lF^{lm}N_m\right)^2,
\end{equation}
and hence conditions \eqref{NEC_FOUT_1} now reads as
\begin{equation} \label{NEC_FOUT}
\renewcommand{\arraystretch}{2.0}
\begin{dcases}
&\tilde{\sigma}^{ij} N_i N_j=\sigma^{ij}N_iN_j-\frac{k}{2}\left(U_lF^{lm}N_m\right)^2<0,
\\
&\sigma^{ij}N_i N_j>0.
\end{dcases}\\[2em]
\end{equation}
Since $\left(U_lF^{lm}N_m\right)^2$ is a positive quantity, the above conditions are both satisfied if
\begin{equation} \label{No_Exotic_Proca}
0<\sigma^{ij}N_iN_j<\frac{k}{2}\left(U_lF^{lm}N_m\right)^2.
\end{equation}
Proca fields satisfying \eqref{No_Exotic_Proca} are possible non-exotic sources for wormholes metric. In other words, for this kind of matter both flaring-out and NEC hold.

\subsection{Rarita-Schwinger Field $(S=\frac{3}{2})$}

 In the theory of supergravity the Rarita-Schwinger field is associated with the conjectural gravitino, the supersymmetric partner of graviton \cite{Watanabe:2004nt,Hayashi:2003cg,Deser:1976eh,ChoquetBruhat:1985iva}. In general, the Rarita-Schwinger Lagrangian  describes a massive particle with spin$-\frac{3}{2}$  \cite{Watanabe:2004nt,ChoquetBruhat:1985iva}
 \begin{eqnarray}
&& \mathcal{L}_{{\rm RS}}= -im \bar{\psi}_i \gamma^{i j} \psi_j
 \nonumber\\ &&-\frac{1}{2}\epsilon^{ijkl} \left[ \overline{\psi} \gamma_5 \gamma_j D_k \psi_l-\overline{\left( D_k \psi_i\right)}\gamma_5\gamma_j \psi_l \right],\,\,\,\,\,\,\,\,
 \end{eqnarray}
 where $\psi_i$ represents a spinor valued one-form and the bar stands for the Dirac conjugate. The covariant derivative of Rarita-Schwinger field is given by
 \begin{equation}
 D_i \psi_j=\left(\partial _i  -\frac{i}{4}\,\omega _i ^{\; \alpha \beta}\, \gamma_{\alpha \beta}\right)\psi_j - \Gamma^k_{\; ji}\psi_k.
 \end{equation}
Unlike the Dirac case, the spin density tensor  is not completely antisymmetric and reads as
\begin{eqnarray}
\tau_{ijk}&=&\frac{i}{4} \left[\bar{\psi}_j\gamma_i \psi_k-\bar{\psi}_k\gamma_i \psi_j \right. \nonumber\\
&+& \eta_{ij} \left(\bar{\psi}_k\gamma^l \psi_l-\bar{\psi}_l\gamma^l \psi_k \right)  \nonumber\\
&-& \left. \eta_{ik}\left(\bar{\psi}_j\gamma^l \psi_l-\bar{\psi}_l\gamma^l\psi_j \right)\right].
\label{R-S_tau_tensor}
\end{eqnarray}
From equation \eqref{torsion_def} one gets the torsion tensor
\begin{equation}
S_{ijk}=-i\frac{k}{2}\left(\bar{\psi}_j\gamma_i \psi_k-\bar{\psi}_k\gamma_i \psi_j\right).
\end{equation}
In order to simplify the subsequent calculations we introduce the tensors $J_{ijk}\equiv \left(\bar{\psi}_i\gamma_j \psi_k-\bar{\psi}_k\gamma_j \psi_i\right)$ and $J_i \equiv J^{j}_{\; ji}$. Note that $J_{ijk}$ is antisymmetric in the first and third indexes. From Eqs. (\ref{sigmatilde}) and \eqref{R-S_tau_tensor} we obtain
\begin{eqnarray}
\tilde{\sigma}_{ij}&=&\sigma_{ij} \nonumber\\
&+& k\left[ 4 J^{(kl)}  _{\; \; \; \; \; \; i}J_{j(kl)}-2J_k\ ^l\ _{(i}J^k \ _{j)l}+ 4J^kJ_{(ij)k}\right. \nonumber\\ 
&+&  \left.\frac{1}{2}g_{ij} \left(J^{klm}J_{klm}+2J^{klm}J_{lkm}-4J^kJ_k\right)\right] \nonumber.
\end{eqnarray}
Also this time we split the correction terms in two: those who do multiply $g_{ab}$ and those who do not. Of course only the seconds produce non vanishing contribution to $\tilde{\sigma}_{ij}N^iN^j$ and are relevant for \eqref{NEC_FOUT_1}. Furthermore it can be proved that the $\tilde{\nabla}_k\tau_{(ij)}\ ^k $ is expected to vanish \cite{Nurgalev:1983vc}. In other words, we are interested only in the following correction term:
\begin{eqnarray} \label{true_correction_RS}
&&T_{ij}^{(\rm corr)}N^iN^j :=k\left[ 4 J^{(kl)}  _{\; \; \; \; \; \; i}J_{j(kl)}-2J_k\ ^l\ _{(i}J^k \ _{j)l} \right.\nonumber\\
&&+  \left. 4J^kJ_{(ij)k}+4J_iJ_j+  \right]N^iN^j,
\end{eqnarray}
which is a scalar. In order to simplify calculations we evaluate \eqref{true_correction_RS} in the proper ``hatted''  reference frame introduced in Ref.\cite{Morris:1988cz} and on the special null vector field $N_{\hat{\mu}}=N(1,1,0,0)$. Hereafter the ``hat'' on hatted coordinate  coordinates will be omitted. After some calculations one obtains
\begin{eqnarray} \label{correction_RS_fin}
&&T_{\mu \nu}^{( \rm corr)}N^{\mu}N^{\nu}=\\
&& 2kN^2\left[ \tau^{1}\tau^{1}-\frac{\tau^a \tau_a}{4}+\frac{\varphi}{4}-\left(\frac{\rho}{4}-\frac{\chi^2}{4}\right)^2\right.\nonumber\\
&&+\tau_{a}J^{(0a)1}+J^{i01}J_{01a}+J_{a01}J^{10a} \nonumber\\
&& +J^{0a1}\left(-J_{0a1}+2J_{10a}-J_{01a}\right)\nonumber\\
&&+\frac{1}{2}J_{a10}\tau^a+J^{ab1}\left(4J_{0(ab)}+2J_{1(ab)}\right)+J^{ab0}J_{a1b}\nonumber \\
&&-J^{ab1}\left(J_{a0b}+J_{a1b}\right)+\frac{\lambda}{\sqrt{2}}\left(J^{110}+\frac{\tau^1}{2}\right)\nonumber\\
&&-2J^{(01)a}v_a -J^{11a}\tau_a +J^{11a}v_a+\frac{\tau^av_a}{2}\nonumber \\
&&+\left.\left(\frac{\lambda}{\sqrt{2}}+v_1\right)\left(\frac{\lambda}{\sqrt{2}}+v_1-\tau_1\right)\right]
\end{eqnarray}
where $a,b \in\{1,2,3\}$ are spatial indexes and $0$  denotes temporal components. We also used the composite fields introduced in Ref. \cite{Watanabe:2004nt}:
\begin{eqnarray}
\begin{array}{cc}
J^{a\mu 0}J_{a\mu 0}=\chi^2/4   &  J^{ab0}J_{0ab}=\rho/4 \\  
J^{ab0}J_{a0b}=\varphi/4  &  J^a\ _{a0}=\lambda/2\sqrt{2} \\ 
J^b\ _{ba}=v_a/2  & J^{00a}=-J^{a00}=\tau^a/2
\end{array}
\end{eqnarray}
This time the correction term \eqref{correction_RS_fin} is positive or negative depending on the field configuration. In particular, If \eqref{correction_RS_fin} is positive, there is no way to satisfy \eqref{NEC_FOUT_1}. We conclude that a Rarita-Schwinger field is able to produce exotic free wormhole solutions only if  $T_{\mu \nu}^{(\rm corr)}N^{\mu}N^{\nu}<0$  and \eqref{NEC_FOUT_1} are both satisfied.

\section{Conclusion}
In Einstein-Cartan framework the flaring-out condition does not always imply NEC violation. This allows to obtain, only for some particular values of the spin,  static spherically symmetric traversable wormholes from non-exotic spinning matter. We analysed four different matter fields with different spin and discussed if the they are able to produce wormholes of the above mentioned kind without violating NEC. For scalar $(S=0)$ and Dirac $(S=1/2)$ fields the exoticity turns out to be an unavoidable feature of wormhole solutions. On the other hand, Proca fields $(S=1)$ satisfying  relation \eqref{NEC_FOUT} are able to produce wormholes without violating NEC. Although Rarita-Schwinger fields $(S=3/2)$ are possible source of wormholes which preserve NEC too,  Eq.\eqref{NEC_FOUT_1} certainly implies exoticity when \eqref{correction_RS_fin} is positive.  All the above discussed systems show how the coupling between spin and torsion provides  more physically attainable  sources for  Morris-Thorne wormholes.

\section*{acknowledgments}
E. B. and E. D. G. are grateful to the Dipartimento di Fisica ``Ettore Pancini'' of Federico II University for
hospitality and support. The work of E. B. has been supported by the INFN funding of the NEWREFLECTIONS experiment.
G.M and M.M. acknowledge support by the Instituto Nazionale di Fisica Nucleare I.S. TAsP and the PRIN 2012 ``Theoretical Astroparticle Physics" of the Italian Ministero dell'Istruzione, Universit\`a e Ricerca.

\bibliographystyle{ieeetr} 
\bibliography{references}
 
\end{document}